\documentclass[aps,floatfix,amsmath,nofootinbib,amssymb,superscriptaddress,letterpaper]{revtex4-2}

\usepackage{overpic}
\usepackage{amssymb}
\usepackage{indentfirst}
\usepackage{feynmf}   
\usepackage{slashed}  
\usepackage{cases}
\usepackage{color}
\usepackage{multirow}
\usepackage{epstopdf}
\usepackage{graphicx,color,bm}
\usepackage{booktabs}
\usepackage{float}
\usepackage{amsmath}
\allowdisplaybreaks[4]

\usepackage[colorlinks=true,
            citecolor=green,
            anchorcolor=red,
            menucolor=red,
            linkcolor=red,
            filecolor=red,
            runcolor=red,
            urlcolor=blue]{hyperref}

\begin{document}

\title{Leading-twist gluon transverse momentum dependent distributions from large-momentum effective theory}

\author{Xiupeng Xie}\affiliation{School of Physics, Southeast University, Nanjing
211189, China}

\author{Zhun Lu}
\email{zhunlu@seu.edu.cn}
\affiliation{School of Physics, Southeast University, Nanjing 211189, China}

\begin{abstract}
We present the perturbative calculation of the leading-twist gluon transverse momentum dependent distribution functions (TMDPDFs) at one-loop level within large-momentum effective theory. By employing the basis of circular polarizations, we generalize the gluon-gluon correlator to a matrix $\Phi_{\Lambda \Lambda^\prime}^{ij}(x,\bm{k}_T^2;S)$ in the gluon $\otimes$ hadron spin space. Incorporating the transverse momentum of external states, we derive the leading-twist gluon TMDPDFs and quasi-TMDPDFs at one-loop level. Our calculations reveal nonzero distributions for the gluon TMDPDFs $f_{1T}^{\perp g}(x,\bm{k}_T^2)$, $h_{1}^{\perp g}(x,\bm{k}_T^2)$, $h_{1T}^{g}(x,\bm{k}_T^2)$, and $h_{1T}^{\perp g}(x,\bm{k}_T^2)$ . These findings enable future lattice QCD simulations to extract these gluon TMDPDFs, which will facilitate a complete determination of nucleon tomography.

\end{abstract}

\maketitle

\section{Introduction}

Fifty years of experimental investigations into the nucleon's internal structure have yielded profound insights into the dynamics of quarks and gluons~\cite{Brambilla:2014jmp}. A cornerstone of Quantum Chromodynamics (QCD) is color confinement, which dictates that quarks and gluons are irrevocably bound within color-singlet hadrons~\cite{tHooft:1973alw,Wilson:1974sk}. This fundamental aspect of the strong interaction precludes the direct observation of isolated partons, implying that the internal structure of hadrons must be inferred from their interactions in high-energy processes. The theoretical framework for making quantitative predictions in such processes is provided by QCD factorization~\cite{Collins:1989gx}. This powerful formalism, underpinned by the property of asymptotic freedom~\cite{Coleman:1973sx,Gross:2005kv,Gross:1973id,Politzer:1973fx}, allows for a systematic separation of perturbatively calculable short-distance dynamics from the non-perturbative, universal parton distribution functions (PDFs) and fragmentation functions (FFs) from perturbatively calculable partonic hard scattering processes.

Early experimental and theoretical efforts primarily investigated the longitudinal momentum structure of quarks and gluons, rather than their confined spatial distribution and motion inside the nucleon. These studies provided a one-dimensional momentum picture, encoded in universal PDFs within the well-established collinear QCD factorization formalism~\cite{Collins:1989gx}. 
In recent years, theoretical advances have extended this description to include transverse dimensions, leading to transverse-momentum-dependent PDFs (TMDPDFs). TMDPDFs play a crucial role in elucidating the three-dimensional structure of hadrons~\cite{Collins:2011zzd,Angeles-Martinez:2015sea,Boer:2011fh,Accardi:2012qut} and are essential for describing high-energy processes~\cite{EuropeanMuon:1991sne,ZEUS:1995acw,H1:1996muf,H1:2008rkk,HERMES:2012uyd,COMPASS:2013bfs,COMPASS:2017mvk}.
such as deeply inelastic scattering, Drell-Yan and hadron pair production in $e^+ e^- $ annihilation.

Lattice Quantum Chromodynamics (LQCD) provides a systematic, first-principles approach for computing non-perturbative QCD observables directly from the QCD Lagrangian. Early exploration of transverse-momentum-dependent structure employed the Lorentz invariance method, yielding ratios of $x$-moments for quark TMDPDFs in pioneering studies~\cite{Musch:2010ka,Musch:2011er,Engelhardt:2015xja,Yoon:2015ocs,Yoon:2017qzo}. 
More recently, the large-momentum effective theory (LaMET) has been established as a powerful framework for extracting light-cone distributions from Euclidean lattice correlators~\cite{Ji:2013dva,Ji:2014gla,Ji:2020ect}. The central strategy of LaMET involves constructing corresponding quasi-distributions. 
These quasi-TMDPDFs are defined as equal-time Euclidean correlators with a finite-length staple-shaped Wilson line of finite length $L$ extending along a spatial direction. 
A factorization theorem, applicable in the large-momentum limit $P^z \to \infty$, sysmetically relates these computable quasi-distributions to their physical light-cone counterparts. 
Thos approach has been successfully applied to compute a wide range of  PDFs~\cite{Lin:2014zya,Alexandrou:2015rja,Chen:2016utp,Alexandrou:2016jqi,Chen:2017mzz,Lin:2017ani,Alexandrou:2017dzj}, meson distribution amplitudes~\cite{Zhang:2017bzy,Zhang:2017zfe}, and quark TMDPDFs~\cite{LatticeParton:2020uhz,Li:2021wvl,LatticePartonLPC:2022eev,Shanahan:2020zxr,Shanahan:2021tst,Schlemmer:2021aij}. In contrast, first-principle calculation of gluon TMDPDFs remains largely unexplored within this framework.

In this work, we present a perturbative calculation of the complete set of leading-twist gluon TMDPDFs and their quasi-distribution counterparts at one-loop order. This calculation establishes a framework for their future extraction from lattice QCD simulations via the LaMET approach. The unpolarized and helicity gluon TMDPDFs, $f_{1}^{g}(x,\bm{k}_T^2)$ and $g_{1L}^{g}(x,\bm{k}_T^2)$, have previously been computed at one-loop level~\cite{Zhu:2022bja}; we note that their determination does not require explicit dependence on the external transverse momentum.

To enable the calculation of the remaining six gluon TMDPDFs, we implement two key techniques. First, we adopt a specific calculational frame where the transverse momentum components are set equal: $\bm{k}_T^x = \bm{k}_T^y $~\cite{Dominguez:2011br,Kovchegov:2025gcg}. Second, by utilizing circular polarization states, we formulate the gluon-gluon correlator as a matrix, $\Phi_{\Lambda \Lambda'}^{ij}(x,\bm{k}_T^2;S)$, in the combined gluon $\otimes$ hadron spin space. 
Our analysis shows that, among these remaining distributions, four receive non-vanishing one-loop contributions: $f_{1T}^{\perp g}(x,\bm{k}_T^2)$, $h_{1}^{\perp g}(x,\bm{k}_T^2)$, $h_{1T}^{g}(x,\bm{k}_T^2)$, and $h_{1T}^{\perp g}(x,\bm{k}_T^2)$. Specifically, we find that $f_{1T}^{\perp g}$ and $h_{1}^{\perp g}$ exhibit a particularly simple structure, receiving contributions from only a single diagram class. In contrast, the expressions for $h_{1T}^{g}$ and $h_{1T}^{\perp g}$ are more involved and share structural similarities with the unpolarized gluon distribution $f_{1}^{g}$.

The remainder of this paper is organized as follows. 
In Section~\ref{section2}, we establish our notation and present a detailed derivation of the one-loop results for the leading-twist gluon TMDPDFs. Section~\ref{section3} is devoted to the parallel calculation of the gluon quasi-TMDPDFs at one-loop order, culminating in the presentation of the factorization and matching relation that connects the renormalized quasi-distributions to their light-cone counterparts. 
Finally, we provide a summary and discussion of our results in Section~\ref{section4}.

\section{Gluon TMDPDFs at one-loop}\label{section2}

The hadronic tensor can be systematically organized into leading and non-leading contributions by employing a suitable parametrization of the hadron momentum and spin vectors via a light-cone decomposition in which two light-like vectors, $n_-$ and $n_+$, satisfying $n_+ \cdot n_- = 1$ and $n_\pm^2 = 0$ can be introduced. An arbitrary vector $a^\mu$ can then be decomposed in light-cone coordinates as $[a^-, a^+, \bm{a}_T]$ or, equivalently, as $a^\mu = a^{-}n_{-}^\mu + a^{+}n_{+}^\mu + a_T^\mu$, where $\bm{a}_T$ is the transverse component.
In a frame where the hadron has no transverse momentum, its four-momentum $P^\mu$ and covariant spin vector $S^\mu$ can be expressed as:
\begin{align}
P & = P^{+} n_{+}+\frac{M^{2}}{2 P^{+}} n_{-}\,,\\
k&=x P^{+} n_{+}+\frac{k^2 + \bm{k}_T^2}{2 x P^{+}} n_{-}+\bm{k}_T\,,\\
S & = S_{ L} \frac{P^{+}}{M} n_{+}-S_{L} \frac{M}{2 P^{+}} n_{-}+\bm{S}_{T}\,,
\end{align}
where $M$ is the hadron mass, $x = k^{+} / P^{+}$ is the longitudinal momentum fraction carried by the gluon, and $\bm{k}_T$ is the  transverse momentum of the gluon. The hadron's polarization of the hadron is described by the longitudinal component $S_L$ and the transverse vector $\bm{S}_{T}$.

The gauge-lnvariant non-local gluon-gluon correlator is given by~\cite{Mulders:2000sh}
\begin{align}
\Phi^{\mu \nu,\rho \sigma}(x,\bm{k}_{T};S)=\frac{1}{xP^+} \int \frac{\mathrm{d}\xi^- \mathrm{d}\bm{\xi}_T}{(2 \pi)^3} e^{-i k \cdot \xi} \left\langle P,S \left| F^{\mu \nu}\left(\frac{\xi}{2}\right) \mathcal{W}_{-\infty}\left(\frac{\xi}{2},-\frac{\xi}{2}\right) F^{\rho \sigma}\left(-\frac{\xi}{2}\right) \right| P,S \right\rangle\,,
\end{align}
where $\mathcal{W}_{-\infty}$ denotes the gauge-link (Wilson line) operator, chosen to be past-pointing. The analysis for the future-pointing link proceeds analogously. The gauge-link operator $\mathcal{W}_{\pm \infty}$ is explicitly given by
\begin{align}
\mathcal{W}_{\pm \infty}(b,a)&=\mathcal{W}_{n_-}^{\dagger}(b,\pm \infty) \mathcal{W}_{T}^{\dagger}(\pm \infty n_-^\mu ;\bm{b}_T^\mu ,\bm{a}_T^\mu) \mathcal{W}_{n_-}(a,\pm \infty)\,,\\
\mathcal{W}_{n_-}(a,\pm \infty)&=\mathcal{P} \mathrm{exp} \left[\mp i g \int_0^{\infty} \mathrm{d}s n_- \cdot A(a^\mu + s n_-^\mu)\right]\,,\\
\mathcal{W}_{T}(y^\mu ;\bm{b}_T^\mu ,\bm{a}_T^\mu)&=\mathcal{P} \mathrm{exp} \left[- i g \int_{\bm{a}_T}^{\bm{b}_T} \mathrm{d}\bm{s}_T \cdot \bm{A}_T (y^\mu + \bm{s}_T^\mu)\right]\,.
\end{align}

We focus on the specific component $\Phi^{\mu \nu,\rho \sigma}(x,\bm{k}_{T};S)$ with $\mu=+$ and $\rho=+$, denoted as $\Phi^{+i ,+j}(x,\bm{k}_{T};S) \equiv \Phi^{ij}(x,\bm{k}_{T};S)$, where $i,j$ are transverse spatial indices. 
This correlator can be decomposed to define the eight leading-twist gluon TMDPDFs through the following projections~\cite{Mulders:2000sh,Meissner:2007rx}:
\begin{align}
\delta_{T}^{i j} \Phi^{i j}\left(x, \bm{k}_{T};S\right)=& f_{1}^{ g}\left(x, \bm{k}_{T}^2\right)-\frac{\varepsilon_{T}^{i j} k_{T}^{i} S_{ T}^{j}}{M} f_{1 T}^{\perp  g}\left(x, \bm{k}_{T}^2\right)\,,\label{eq:df} \\
i \varepsilon_{T}^{i j} \Phi^{i j}\left(x, \bm{k}_{T};S\right)=& \lambda g_{1 L}^{g}\left(x, \bm{k}_{T}^2\right)+\frac{\bm{k}_{T} \cdot \bm{S}_{ T}}{M} g_{1 T}^{ g}\left(x, \bm{k}_{T}^2\right)\,, \label{eq:dg} \\
-\hat{S} \Phi^{ i j}\left(x, \bm{k}_{T};S\right)=& -\frac{\hat{S} k_{T}^{i} k_{T}^{j} }{2 M^{2}} h_{1}^{\perp g}\left(x, \bm{k}_{T}^2\right)+\frac{\hat{S} k_{T}^{i} \varepsilon_{T}^{j k} S_{T}^{k}}{2 M} h_{1T}^{ g}\left(x, \bm{k}_{T}^2\right) \notag \\
&+\frac{\hat{S} k_{T}^{i} \varepsilon_{T}^{j k} k_{T}^{k}}{2 M^2}\left(\lambda h_{1 L}^{\perp g}\left(x, \bm{k}_{T}^2\right)+\frac{\bm{k}_{T} \cdot \bm{S}_{ T}}{M} h_{1 T}^{\perp g}\left(x, \bm{k}_{T}^2\right)\right)\,.\label{eq:dh}
\end{align}
In this expression, we employ the systematic transverse tensor $\delta_T^{ij}=-g_T^{ij}$, where $g_T^{ij}=g^{ij}-n_+^i n_-^j -n_+^j n_-^i$, and the anti-systematic transverse tensor $\varepsilon_T^{ij}$ defined such as  $\varepsilon_T^{12}=1$. The symmetrization operator $\hat{S}$ on a generic tensor $O^{ij}$ is defined as:
\begin{align}
\hat{S} O^{ij}=\frac{1}{2}\left(O^{ij}+O^{ji}-\delta_T^{ij} O^{kk}\right)\,.
\end{align}

At leading order, only the unpolarized and helicity gluon TMDPDFs, $f_{1}^{g}(x, \bm{k}_{T}^2)$ and $g_{1L}^{g}(x, \bm{k}_{T}^2)$, can be calculated without explicit dependence on the gluon transverse momentum $\bm{k}_T$ or the hadron mass $M$. In contrast, the computation of the remaining six TMDPDFs requires careful treatment of these parameters. To facilitate this calculation, it is advantageous to express the correlator $\Phi^{ij}$ as a matrix in the combined gluon $\otimes$ hadron spin space~\cite{Mulders:2000sh}:
\begin{align}
\Phi^{ij}=\frac{1}{2}\left\{\Phi_{++}^{ij}+\Phi_{--}^{ij}\right\}+\frac{S_L}{2}\left\{\Phi_{++}^{ij}-\Phi_{--}^{ij}\right\}+\frac{S_T^1}{2}\left\{\Phi_{+-}^{ij}+\Phi_{-+}^{ij}\right\}+\frac{i S_T^2}{2}\left\{\Phi_{+-}^{ij}-\Phi_{-+}^{ij}\right\}\,.
\end{align}
Here, the subscripts $\Lambda, \Lambda\prime = \pm$ denote the helicity states of the incoming and outgoing gluons in the correlator.

Using the circular polarization vectors defined by $\varepsilon_\pm^j\equiv \left[\mp e_x^j-i e_y^j\right]/\sqrt{2}$ with $e_x^i=(1,0)$ and $e_y^i=(0,1)$ in the transverse plane, we construct four essential projection operators for the matrix elements $\Phi_{\Lambda \Lambda^\prime}^{ij}$:
\begin{align}
\frac{1}{2}\left(\varepsilon_+^m \varepsilon_+^{*n}+\varepsilon_-^m \varepsilon_-^{*n}\right)&=\frac{1}{2}\left(\begin{matrix}1&0\\0&1\\\end{matrix}\right)=\frac{1}{2}\left(e_x^m e_x^{n}+e_y^m e_y^{n}\right)=-\frac{g_T^{mn}}{2}\,,\label{ep1}\\
\frac{1}{2}\left(\varepsilon_+^m \varepsilon_+^{*n}-\varepsilon_-^m \varepsilon_-^{*n}\right)&=-\frac{1}{2}\left(\begin{matrix}0&i\\-i&0\\\end{matrix}\right)=\frac{i}{2}\left(e_y^m e_x^{n}-e_x^m e_y^{n}\right)=-\frac{i \varepsilon_T^{mn}}{2}\,,\label{ep2}\\
\frac{S_T^1}{2}\left(\varepsilon_+^m \varepsilon_-^{*n}+\varepsilon_-^m \varepsilon_+^{*n}\right)&=\frac{S_T^1}{2}\left(\begin{matrix}-1&0\\0&1\\\end{matrix}\right)=\frac{S_T^1}{2}\left(e_y^m e_y^{n}-e_x^m e_x^{n}\right)\,,\label{ep3}\\
\frac{iS_T^2}{2}\left(\varepsilon_+^m \varepsilon_-^{*n}-\varepsilon_-^m \varepsilon_+^{*n}\right)&=\frac{S_T^2}{2}\left(\begin{matrix}0&1\\1&0\\\end{matrix}\right)=\frac{S_T^2}{2}\left(e_x^m e_y^{n}+e_y^m e_x^{n}\right)\,.\label{ep4}
\end{align}

To simplify the tensor projections that define the TMDPDFs, we adopt a specific frame where the two components of the gluon transverse momentum are set equal: $\bm{k}_T^x = \bm{k}_T^y = M$~\cite{Dominguez:2011br,Kovchegov:2025gcg}. In this frame, the projection operators appearing in Eq.~(\ref{eq:dh}) simplify to the following matrix forms:
\begin{align}
\frac{\hat{S} k_{T}^{i} k_{T}^{j} }{2 M^{2}}&=\frac{1}{2}\left(\begin{matrix}0&1\\1&0\\\end{matrix}\right)=\frac{1}{2}\left(e_x^i e_y^{j}+e_y^i e_x^{j}\right)=\frac{1}{2i}\left(\varepsilon_+^{*i} \varepsilon_-^{j}-\varepsilon_-^{*i} \varepsilon_+^{j}\right)\,,\\
\frac{\hat{S} k_{T}^{i} \varepsilon_{T}^{j k} k_{T}^{k}}{2 M^2}&=\frac{1}{2}\left(\begin{matrix}1&0\\0&-1\\\end{matrix}\right)=\frac{1}{2}\left(e_x^i e_x^{j}-e_y^i e_y^{j}\right)=-\frac{1}{2}\left(\varepsilon_+^{*i} \varepsilon_-^{j}+\varepsilon_-^{*i} \varepsilon_+^{j}\right)\,,\\
\frac{\hat{S} k_{T}^{i} \varepsilon_{T}^{j k} S_{T}^{k}}{2 M}&=\frac{S_T^1}{4}\left(\begin{matrix}1&-1\\-1&-1\\\end{matrix}\right)+\frac{S_T^2}{4}\left(\begin{matrix}1&1\\1&-1\\\end{matrix}\right)\,.
\end{align}

\begin{figure}
    \centering
    \includegraphics[width=0.6\columnwidth]{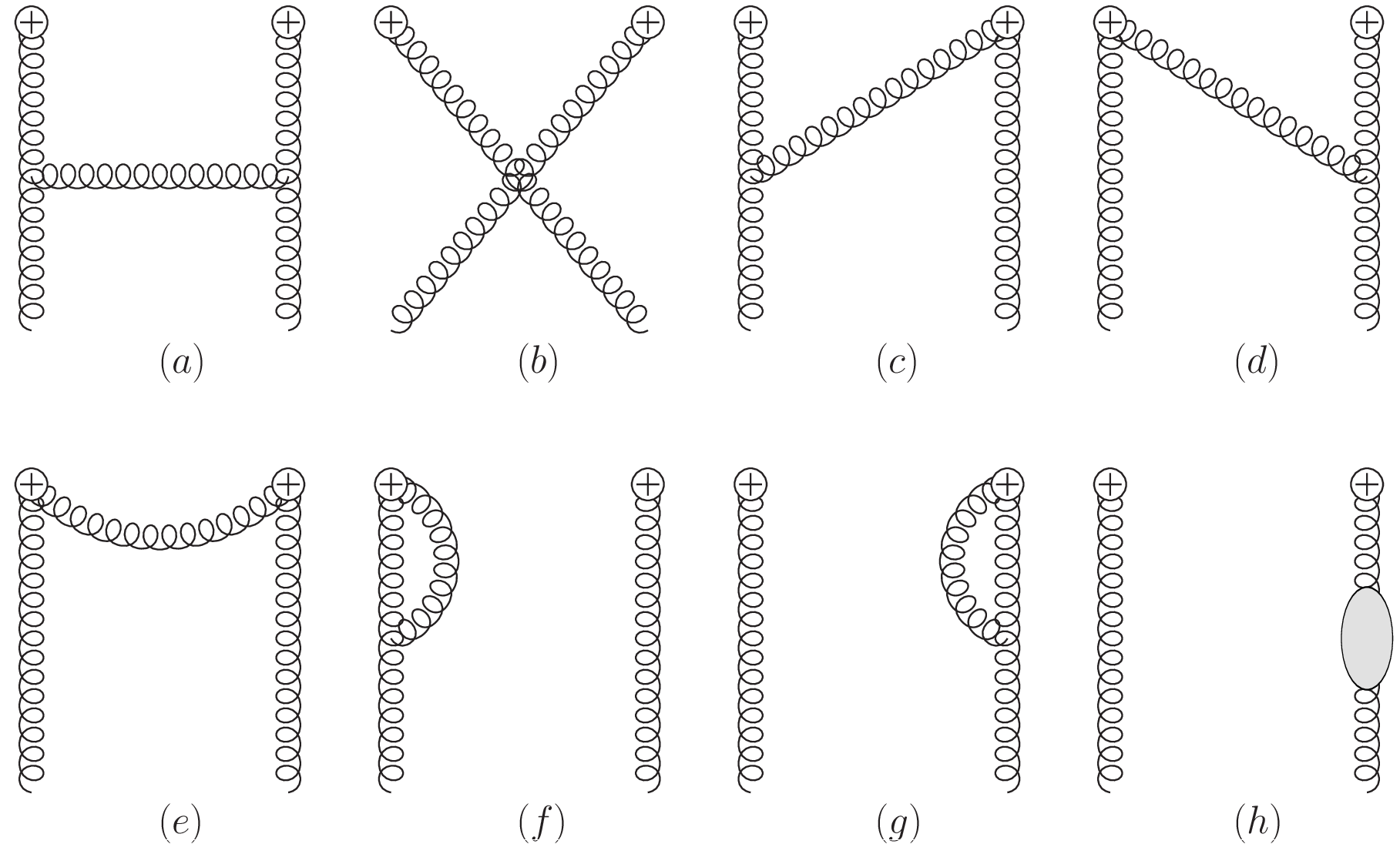}
    \caption{One-loop Feynman diagrams contributing to the gluon TMDPDFs without interactions with the Wilson line.}
    \label{oneloop1}
\end{figure}
\begin{figure}
    \centering
    \includegraphics[width=0.6\columnwidth]{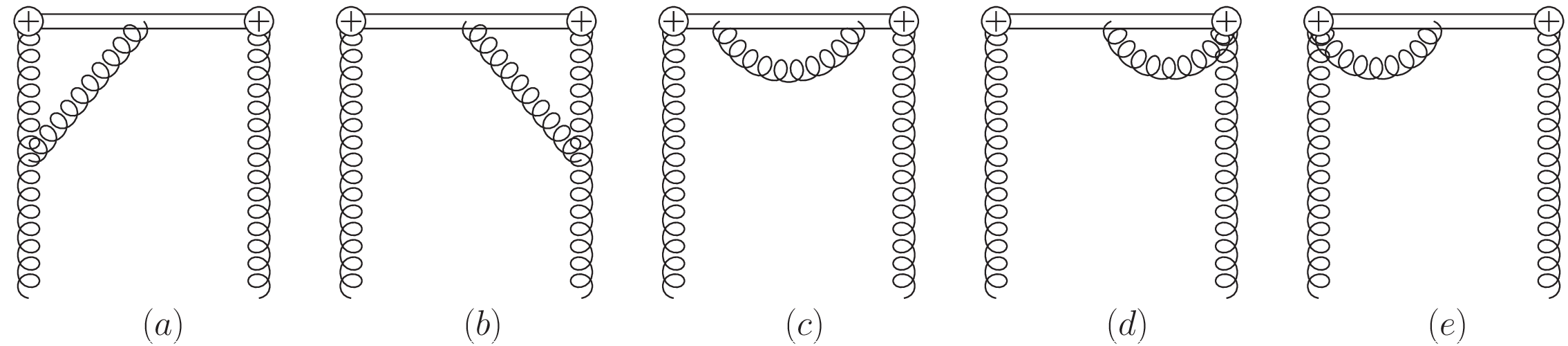}
    \caption{One-loop Feynman diagrams contributing to the gluon TMDPDFs involving interactions with the Wilson line.}
    \label{oneloop2}
\end{figure}

For a perturbative calculation of the unsubtracted gluon-in-gluon TMDPDFs, the hadronic state $|P,S\rangle$ is replaced by an on-shell gluon state $\left.\right|p,\varepsilon\rangle$, where $p$ is the gluon momentum and $\varepsilon$ is the gluon polarization vector, satisfying the on-shell conditions $p^2=0$ and $p\cdot\varepsilon=0$. We work in Feynman gauge and categorize the one-loop corrections to the unsubtracted gluon TMDPDFs into two catalogs: diagrams without gauge-link interactions, shown in Fig.~\ref{oneloop1}; and diagrams that contain such interactions, shown in Fig.~\ref{oneloop2}. Both ultraviolet (UV) and infrared (IR) divergences are regulated using dimensional regularization in $D = 4 - 2\epsilon$ dimensions.

The contribution to the gluon-gluon correlator $\Phi^{ij}$ from the diagrams in Fig.~\ref{oneloop1} is given by
\begin{align}
&\left.\Phi^{i j}\left(x, \bm{k}_{T};S\right)\right|_{\ref{oneloop1} (a)}\notag\\
&=\frac{1}{xP^+} \int \frac{\mathrm{d}\xi^- \mathrm{d}\bm{\xi}_T}{(2 \pi)^3} e^{-i k \cdot \xi}\left.\left\langle p, \varepsilon^*\left|F^{+i}\left(\frac{\xi}{2}\right) F^{+j}\left(-\frac{\xi}{2}\right)\right|p, \varepsilon\right\rangle\right|_{\ref{oneloop1} (a)}\notag\\
&= \frac{1}{xP^+} \int \frac{\mathrm{d}\xi^- \mathrm{d}\bm{\xi}_T}{(2 \pi)^3} e^{-i k \cdot \xi}  \tilde{\mu}^{2\epsilon} \int \frac{\mathrm{d}^D k}{(2 \pi)^D} e^{i k\cdot \xi} \notag \\
&\phantom{{}={}} \times (-g f_{a_1 b_1 c_1})\left[(p+k)^{\gamma_1}g^{\mu_1 \nu_1}+(-2k+p)^{\mu_1}g^{\nu_1 \gamma_1}+(k-2p)^{\nu_1}g^{\mu_1 \gamma_1}\right] \delta_{b_1 b_2}\notag\\
&\phantom{{}={}} \times (-g f_{a_2 c_1 b_2})\left[(p+k)^{\gamma_2}g^{\mu_2 \nu_2}+(-2k+p)^{\mu_2}g^{\nu_2 \gamma_2}+(k-2p)^{\nu_2}g^{\mu_2 \gamma_2}\right]\notag\\
&\phantom{{}={}} \times (-i) \left(k_+ g^{\nu_1^\prime i}-k^i n_-^{\nu_1^\prime}\right)(i) \left(k_+ g^{\nu_2^\prime j}-k^j n_-^{\nu_2^\prime}\right)\notag\\
&\phantom{{}={}} \times \frac{-i g_{\nu_1 \nu_1^\prime}}{k^2} \frac{-i g_{\nu_2 \nu_2^\prime}}{k^2} \frac{-i g_{\gamma_1 \gamma_2}}{(p-k)^2} \varepsilon_{\mu_1} \varepsilon_{\mu_2}^* \frac{\delta_{a_1 a_2}}{N_c^2 - 1}\,,
\end{align}
where $\tilde{\mu} = \mu \sqrt{e^{\gamma_E}/(4\pi)}$ is the $\overline{\text{MS}}$ renormalization scale parameter and $\gamma_E$ is the Euler-Mascheroni constant. To extract a specific gluon TMDPDF, the gluon polarization sum $\varepsilon_{\mu_1} \varepsilon_{\mu_2}^*$ is replaced by the corresponding projection operator from Eqs.~~(\ref{ep1})-(\ref{ep4}). For instance, the linearly polarized gluon distribution $h_1^{\perp g}$ is obtained via
\begin{align}
\frac{\hat{S} k_{T}^{i} k_{T}^{j} }{2 M^{2}} h_{1}^{\perp g}\left(x, \bm{k}_{T}^2\right)= \left.\hat{S} \Phi^{ i j}\left(x, \bm{k}_{T};S\right)\right|_{\varepsilon_{\mu_1} \varepsilon_{\mu_2}^*\rightarrow -g_{T, \mu_1 \mu_2}/2}\,.
\end{align}

In the diagrams of Fig.~\ref{oneloop2}, the contributions from the gauge-link must be accounted for. To regularize the resulting rapidity divergences, we employ the $\delta$-regularization method~\cite{Echevarria:2011epo,Echevarria:2014rua,Vladimirov:2014aja,
Cherednikov:2008ua,Echevarria:2015usa,Echevarria:2015byo}, which modifies the eikonal propagator associated with the Wilson line as follows:
\begin{align}
\frac{i}{k^{\pm}\pm i0}\rightarrow \frac{i}{k^{\pm}\pm i\delta^\pm}\,.
\end{align}

We now present the expressions for the leading-twist gluon TMDPDFs at one-loop order. Our results reveal several noteworthy features. First, aside from the unpolarized and helicity gluon TMDPDFs $f_{1}^{g}(x,\bm{k}_T^2)$ and $g_{1L}^{g}(x,\bm{k}_T^2)$, whose one-loop results are already known~\cite{Zhu:2022bja}, four of the remaining six gluon TMDPDFs receive non-vanishing one-loop contributions: $f_{1T}^{\perp g}(x,\bm{k}_T^2)$, $h_{1}^{\perp g}(x,\bm{k}_T^2)$, $h_{1T}^g(x,\bm{k}_T^2)$ and $h_{1T}^{\perp g}(x,\bm{k}_T^2)$ exhibit nonzero distributions at one-loop level. In contrast, our perturbative calculations yields null results for $g_{1T}^{ g}(x,\bm{k}_T^2)$ and $h_{1L}^{\perp g}(x,\bm{k}_T^2)$. 
Physically, a vanising result for $g_{1T}^{ g}(x,\bm{k}_T^2)$ indicates zero probability to find a transversely polarized gluon inside a longitudinally polarized gluon target at this order, while the null result for $h_{1L}^{\perp g}(x,\bm{k}_T^2)$ corresponds to zero probability for a longitudinally polarized gluon inside a transversely polarized target. 
Furthermore, the distributions $f_{1T}^{\perp g}(x,\bm{k}_T^2)$ and $h_{1}^{\perp g}(x,\bm{k}_T^2)$ exhibit a particularly simple structure, receiving contributions solely from diagram Fig.~\ref{oneloop1}(a). 
It is important to note that both $f_{1T}^{\perp g}$ and $h_{1}^{\perp g}$ also vanish at tree level. 
In addition, the results for $f_{1T}^{\perp g}$, $h_{1T}^g$, and $h_{1T}^{\perp g}$ depend on the orientation of the transverse spin vector $\bm{S}_T$. In this work, we present results for the two representative orientations: $\bm{S}_T = (S_T^1, 0)$ and $\bm{S}_T = (0, S_T^2)$.

The results of $f_{1T}^{\perp g}(x,\bm{k}_T^2)$ and $h_{1}^{\perp g}(x,\bm{k}_T^2)$ read
\begin{align}
\left.x f_{1T}^{\perp g}(x,\bm{k}_T^2;S_T^1)\right|_{\ref{oneloop1} (a)}&=0\,,\\
\left.x f_{1T}^{\perp g}(x,\bm{k}_T^2;S_T^2)\right|_{\ref{oneloop1} (a)}&=\frac{\alpha_s C_A}{\pi^2 }\frac{(2 \pi \tilde{\mu})^{2\epsilon}}{ \bm{k}_T^2} (x-1)x^2\,,\\
\left.x h_{1}^{\perp g}(x,\bm{k}_T^2)\right|_{\ref{oneloop1} (a)}&=\frac{\alpha_s C_A}{\pi^2 }\frac{(2 \pi \tilde{\mu})^{2\epsilon}}{ \bm{k}_T^2} (1-x)\,.
\end{align}

For $h_{1T}^{\perp g}(x,\bm{k}_T^2)$, the results are identical for both transverse spin orientations, i.e., $h_{1T}^{\perp g}(x,\bm{k}_T^2; S_T^1) = h_{1T}^{\perp g}(x,\bm{k}_T^2; S_T^2)$:
\begin{align}
\left.x h_{1T}^{\perp g}(x,\bm{k}_T^2)\right|_{\ref{oneloop1} (a)}&=\frac{\alpha_s C_A}{2\pi^2 }\frac{(2 \pi \tilde{\mu})^{2\epsilon}}{ \bm{k}_T^2} x^2\,,\label{h1_1a}\\
\left.x h_{1T}^{\perp g}(x,\bm{k}_T^2)\right|_{\ref{oneloop1} (c)}&=\left.x h_{1T}^{\perp g}(x,\bm{k}_T^2)\right|_{\ref{oneloop1} (d)}=-\frac{\alpha_s C_A}{4\pi^2 }\frac{(2 \pi \tilde{\mu})^{2\epsilon}}{ \bm{k}_T^2} x(1+x)\,,\\
\left.x h_{1T}^{\perp g}(x,\bm{k}_T^2)\right|_{\ref{oneloop1} (f)}&=\left.x h_{1T}^{\perp g}(x,\bm{k}_T^2)\right|_{\ref{oneloop1} (g)}=-\frac{3\alpha_s C_A}{8\pi }\delta(1-x) \delta^2(\bm{k}_T^2) \left(\frac{1}{\epsilon_{UV}}-\frac{1}{\epsilon_{IR}}+\ln \frac{\mu_{UV}^2}{\mu_{IR}^2}\right)\,,\\
\left.x h_{1T}^{\perp g}(x,\bm{k}_T^2)\right|_{\ref{oneloop1} (h)+h.c.}&=\frac{\alpha_s}{\pi} \delta(1-x) \delta^2(\bm{k}_T^2) \left(\frac{5}{12}C_A -\frac{1}{3}T_F n_f\right) \left(\frac{1}{\epsilon_{UV}}-\frac{1}{\epsilon_{IR}}+\ln \frac{\mu_{UV}^2}{\mu_{IR}^2}\right)\,,\\
\left.x h_{1T}^{\perp g}(x,\bm{k}_T^2)\right|_{\ref{oneloop2} (a)+\ref{oneloop2} (b)}&=\frac{\alpha_s}{4\pi^2} \frac{(2 \pi \tilde{\mu})^{2\epsilon}}{ \bm{k}_T^2} \left[\left(\frac{x(1+x)}{1-x}\right)_+ + \delta(1-x) \left(-2 \ln \frac{\delta^+}{p^+}+i \pi -\frac{5}{2}\right)\right]\notag\\
&\phantom{{}={}} +\frac{\alpha_s}{4\pi} \delta(1-x) \delta^2(\bm{k}_T^2) \left(\frac{1}{\epsilon_{UV}}-\frac{1}{\epsilon_{IR}}+\ln \frac{\mu_{UV}^2}{\mu_{IR}^2}\right) \left(-2 \ln \frac{\delta^+}{p^+}+i \pi -\frac{5}{2}\right)+h.c.\,.\label{h1_2ab}
\end{align}
Notably, diagrams Fig.~\ref{oneloop1}(b), Fig.~\ref{oneloop1}(e) and Figs.~\ref{oneloop2}(c)-\ref{oneloop2}(e) yield vanishing contributions to this TMDPDF.

In the case of $h_{1T}^g(x,\bm{k}_T^2)$, a non-zero result emerges only for the specific spin orientation corresponding to $h_{1T}^g(x,\bm{k}_T^2; S_T^2)$:
\begin{align}
\left.xh_{1T}^g(x,\bm{k}_T^2;S_T^1)\right|_{\ref{oneloop1} (a)\sim(h)}&=0\,,\\
\left.xh_{1T}^g(x,\bm{k}_T^2;S_T^2)\right|_{\ref{oneloop1} (a)\sim(h)}&=-2 \cdot \left.x h_{1T}^{\perp g}(x,\bm{k}_T^2)\right|_{\ref{oneloop1} (a)\sim(h)}\,.
\end{align}
This distinct behavior motivates the introduction of a unified expression:
\begin{align}
x \Delta H_T^g(x,\bm{k}_T^2)=-\left(xh_{1T}^g(x,\bm{k}_T^2)+\frac{\bm{k}_T^2}{2M^2}x h_{1T}^{\perp g}(x,\bm{k}_T^2)\right)=\kappa x h_{1T}^{\perp g}(x,\bm{k}_T^2)\,.
\end{align}
Here, the coefficient $\kappa = \pm 1$ is introduced to explicitly capture the sign change under different orientations of $\bm{S}_T$: $\kappa = -1$ for $\bm{S}_T = (S_T^1, 0)$, and $\kappa = +1$ for $\bm{S}_T = (0, S_T^2)$.

Combining Eqs.~(\ref{h1_1a})-(\ref{h1_2ab}) and performing the Fourier transform to coordinate space,
\begin{align}
h_{1T}^{\perp g}(x,\bm{b}_T^2)=\int \mathrm{d}^2 \bm{k}_T^2 e^{-i \bm{k}_T \cdot \bm{b}_T} h_{1T}^{\perp g}(x,\bm{k}_T^2)\,,
\end{align}
we obtain the unsubtracted gluon TMDPDF in $(x,\bm{b}_T)$ space 
\begin{align}
x h_{1T}^{\perp g}(x,\bm{b}_T^2)=&x h_{1T}^{\perp g,(0)}(x,\bm{b}_T^2)+x h_{1T}^{\perp g,(1)}(x,\bm{b}_T^2)\notag\\
=&\delta(1-x) -\frac{\alpha_s}{2\pi} \left(\frac{1}{\epsilon_{IR}}+\ln \frac{\mu^2 \bm{b}_T^2 e^{2 \gamma_E}}{4}\right)\cdot x \left[C_A \frac{2x}{(1-x)_+} + \frac{\beta_0}{2 } \delta (1-x)\right]\notag\\
&+\frac{\alpha_s C_A}{2\pi} \left(\frac{1}{\epsilon_{UV}}+\ln \frac{\mu^2 \bm{b}_T^2 e^{2 \gamma_E}}{4}\right) \left(2\ln \frac{\delta^+}{p^+}+\frac{\beta_0}{2 C_A}\right) \delta(1-x)\,,\\
x \Delta H_T^g(x,\bm{b}_T^2)=&\kappa x h_{1T}^{\perp g}(x,\bm{b}_T^2)\,,
\end{align}
where the superscripts (0) and (1) represent the zeroth order and one-loop order results, respectively, $\beta_0=\frac{11 C_A -4 T_F n_f}{3}$  and we have set $\mu_{UV}=\mu_{IR}=\mu$ . 
This result differs from the unpolarized and helicity gluon TMDPDFs, $f_{1}^{g}(x,\bm{b}_T^2)$ and $g_{1L}^{g}(x,\bm{b}_T^2)$, only in the specific form of the gluon-gluon splitting kernel.

To obtain a UV-finite and rapidity-divergence-free quantity, we employ the soft function from Ref.~\cite{Zhu:2022bja},
\begin{align}
S\left(\bm{b}_T,\mu,\delta^+,\delta^-\right)=&1+\frac{\alpha_s C_A}{2\pi}\left(-\frac{2}{\epsilon_{UV}^2}+\frac{2}{\epsilon_{UV}} \ln \frac{2\delta^+ \delta^-}{\mu^2}+ \ln^2 \frac{\mu^2 \bm{b}_T^2 e^{2\gamma_E}}{4}\right.\notag\\
&\left.+2 \ln \frac{\mu^2 \bm{b}_T^2 e^{2\gamma_E}}{4} \ln \frac{2\delta^+ \delta^-}{\mu^2} +\frac{\pi^2}{6}\right)\,,
\end{align}
along with the $\overline{\mathrm{MS}}$ UV renormalization factor,
\begin{align}
Z\left( \mu,\xi,\epsilon \right)=1-\frac{\alpha_s C_A}{2\pi}\left[\frac{1}{\epsilon^2_{UV}}+\frac{1}{\epsilon_{UV}}\left(\frac{\beta_0}{2C_A}+\ln \frac{\mu^2}{\xi}\right)\right]\,.
\end{align}
The renormalized and subtracted gluon TMDPDF is then given by
\begin{align}
x h_{1T,sub}^{\perp g}(x,\bm{b}_T^2,\mu,\xi)=&\frac{x h_{1T}^{\perp g}(x,\bm{b}_T^2) Z\left( \mu,\xi,\epsilon \right) }{\sqrt{S\left(\bm{b}_T,\mu,\delta^+,\delta^-\right)}} \notag\\
=&\delta(1-x) -\frac{\alpha_s}{2\pi} \left(\frac{1}{\epsilon_{IR}}+\ln \frac{\mu^2 \bm{b}_T^2 e^{2 \gamma_E}}{4}\right)\cdot x \left[C_A \frac{2x}{(1-x)_+} + \frac{\beta_0}{2 } \delta (1-x)\right]\notag\\
&+\frac{\alpha_s C_A}{2\pi} \left[\ln \frac{\mu^2 \bm{b}_T^2 e^{2\gamma_E}}{4} \left(\frac{\beta_0}{2C_A}+\ln \frac{\mu^2}{\xi}\right)-\frac{1}{2}\ln^2 \frac{\mu^2 \bm{b}_T^2 e^{2\gamma_E}}{4} -\frac{\pi^2}{12} \right] \delta(1-x)\,,
\end{align}
where $\delta^+=\delta^-$ and $\xi=2(p^+)^2=2(xP^+)^2$. As expected, the subtracted gluon TMDPDF $\Delta H_{T,sub}^{g}(x,\bm{b}_T^2)$ maintains the same relation with $h_{1T,sub}^{\perp g}(x,\bm{b}_T^2,\mu,\xi)$, differing only by the overall factor $\kappa$.

\section{Gluon quasi-TMDPDFs at one-loop}\label{section3}

The gluon quasi-TMDPDFs are defined through the equal-time non-local correlator
\begin{align}
\tilde{\Phi}^{\mu \nu,\rho \sigma}(x,\bm{k}_{T},S,P^z)=\lim_{L\rightarrow \infty} \frac{N}{x} \int \frac{\mathrm{d}\eta^z \mathrm{d}\bm{\eta}_T}{(2 \pi)^3} e^{-i k \cdot \eta} \frac{\left\langle P,S \left| F^{\mu \nu}\left(\frac{\eta}{2}\right) \widetilde{\mathcal{W}}_{-L}\left(\frac{\eta}{2},-\frac{\eta}{2}\right) F^{\rho \sigma}\left(-\frac{\eta}{2}\right) \right| P,S \right\rangle}{\sqrt{Z_E\left(2L, \left|\bm{b}_T\right|\right)}}\,,\label{quasiphi}
\end{align}
where the hadron momentum is chosen as $P^\mu=(P^0,\bm{0}_T,P^z)$, and the space-like separation is $\eta^\mu=(0,\eta^z,\bm{\eta}_T)$ with $\eta^z = b^z$ and $\bm{\eta}_T = \bm{b}_T$. The staple-shaped gauge links with finite extent $L$ along the $z$-direction are defined by:
\begin{align}
\widetilde{\mathcal{W}}_{\pm L}(b,a)&=\widetilde{\mathcal{W}}_{n_z}^{\dagger}(b,\pm L) \widetilde{\mathcal{W}}_{T}^{\dagger}(\pm L n_z^\mu ;\bm{b}_T^\mu ,\bm{a}_T^\mu) \widetilde{\mathcal{W}}_{n_z}(a,\pm L)\,,\\
\widetilde{\mathcal{W}}_{n_z}(a,\pm \infty)&=\mathcal{P} \mathrm{exp} \left[\mp i g \int_0^{L} \mathrm{d}s n_z \cdot A(a^\mu + s n_z^\mu)\right]\,,\\
\widetilde{\mathcal{W}}_{T}(y^\mu ;\bm{b}_T^\mu ,\bm{a}_T^\mu)&=\mathcal{P} \mathrm{exp} \left[- i g \int_{\bm{a}_T}^{\bm{b}_T} \mathrm{d}\bm{s}_T \cdot \bm{A}_T (y^\mu + \bm{s}_T^\mu)\right]\,.
\end{align}
The space-like and time-like vectors are introduced as
\begin{align}
n_0^\mu&=(1,0,0,0)\,,\quad &n_z^\mu=(0,0,0,-1)\,,\\
n_x^\mu&=(0,-1,0,0)\,,\quad &n_y^\mu=(0,0,-1,0)\,.
\end{align}

The dependence on the length $L$ is cancelled by the square root of the Euclidean flat rectangular Wilson loop in the denominator:
\begin{align}
Z_E\left(2L, \left|\bm{b}_T\right|,\mu\right)=\frac{1}{N_c^2-1} \left\langle 0 \left| \widetilde{\mathcal{W}}_{-2L}\left(\frac{\bm{b}_T^\mu}{2}+L n_z^\mu,-\frac{\bm{b}_T^\mu}{2}+L n_z^\mu\right) \mathcal{W}_{T}\left(L n_z^\mu;-\frac{\bm{b}_T^\mu}{2}, \frac{\bm{b}_T^\mu}{2}\right) \right| 0 \right\rangle\,,\label{quasiZE}
\end{align}
which renders the $L \rightarrow \infty$ limit well-defined.

In this work, we calculate the gluon quasi-TMDPDFs using the specific correlator component $\tilde{\Phi}^{0i,0j} \equiv \tilde{\Phi}^{ij}$, where $i,j$ are transverse spatial indices. This choice corresponds to one of the multiplicatively renormalizable operators identified in Ref.~\cite{Zhang:2018diq}. The normalization factor $N$ in Eq.~(\ref{quasiphi}) is $N = n_z \cdot P / (n_0 \cdot P)^2 = P^z / (P^0)^2$. For an on-shell gluon state $\left.\right|p,\varepsilon\rangle$ with momentum $p^{\mu} = (p^z,\bm{0}_T,p^z)$ and $p^z=x P^z$, this simplifies to $N = 1/p^z$. The leading-twist gluon quasi-TMDPDFs are then defined through the projections:
\begin{align}
\delta_{T}^{i j} \tilde{\Phi}^{i j}\left(x, \bm{k}_{T};S,p^z\right)=& \tilde{f}_{1}^{ g}\left(x, \bm{k}_{T}^2,p^z\right)-\frac{\varepsilon_{T}^{i j} k_{T}^{i} S_{ T}^{j}}{M} \tilde{f}_{1 T}^{\perp  g}\left(x, \bm{k}_{T}^2,p^z\right)\,,\label{eq:df-quasi} \\
i \varepsilon_{T}^{i j} \tilde{\Phi}^{i j}\left(x, \bm{k}_{T};S,p^z\right)=& \lambda \tilde{g}_{1 L}^{g}\left(x, \bm{k}_{T}^2,p^z\right)+\frac{\bm{k}_{T} \cdot \bm{S}_{ T}}{M} \tilde{g}_{1 T}^{ g}\left(x, \bm{k}_{T}^2,p^z\right)\,, \label{eq:dg-quasi} \\
-\hat{S} \tilde{\Phi}^{ i j}\left(x, \bm{k}_{T};S,p^z\right)=& -\frac{\hat{S} k_{T}^{i} k_{T}^{j} }{2 M^{2}} \tilde{h}_{1}^{\perp g}\left(x, \bm{k}_{T}^2,p^z\right)+\frac{\hat{S} k_{T}^{i} \varepsilon_{T}^{j k} S_{T}^{k}}{2 M} \tilde{h}_{1T}^{ g}\left(x, \bm{k}_{T}^2,p^z\right) \notag \\
&+\frac{\hat{S} k_{T}^{i} \varepsilon_{T}^{j k} k_{T}^{k}}{2 M^2}\left(\lambda \tilde{h}_{1 L}^{\perp g}\left(x, \bm{k}_{T}^2,p^z\right)+\frac{\bm{k}_{T} \cdot \bm{S}_{ T}}{M} \tilde{h}_{1 T}^{\perp g}\left(x, \bm{k}_{T}^2,p^z\right)\right)\,.\label{eq:dh-quasi}
\end{align}

For the diagrams in Fig.~\ref{oneloop1}, the results for the gluon quasi-TMDPDFs concide with those for the standard gluon TMDPDFs at the leading power:
\begin{align}
x \tilde{F}^{g}\left(x,\bm{b}_T^2,p^z\right)_{\ref{oneloop1} (a)\sim(h)}=x F^{g}\left(x,\bm{b}_T^2\right)_{\ref{oneloop1} (a)\sim(h)}\,,
\end{align}
where $\left\{F\right\}=\left\{f_{1T}^{\perp},g_{1T},h_{1}^{\perp},h_{1L}^{\perp},h_{1T}^{\perp},h_{1T}\right\}$. Other contributions to the quasi-TMDPDFs are suppressed in the limit $b^z \ll \bm{b}_T \ll L$. Consequently, the gluon quasi-TMDPDFs exhibit physical support $x \in [0,1]$ at leading power.

For the diagrams in Fig.~\ref{oneloop2} involving gauge-link interactions, we calculate the quasi-TMDPDFs following the method outlined for the quark case in Appendix C of Ref.~\cite{Ebert:2019okf}. The sail diagrams (Figs.~\ref{oneloop2}(a) and (b)) contribute to $x \tilde{h}_{1T}^{\perp g}(x,\bm{b}_T^2,p^z)$ as:
\begin{align}
x \tilde{h}_{1T}^{\perp g}(x,\bm{b}_T^2,p^z)|_{\ref{oneloop2} (a)+\ref{oneloop2} (b)}=&\frac{\alpha_s C_A}{2\pi} \left\{-\left(\frac{1}{\epsilon_{IR}}+\ln \frac{\mu^2_{IR} \bm{b}_T^2 e^{2\gamma_E}}{4}\right) \left[\frac{x(1+x)}{1-x}\right]_+ +\delta(1-x)\left[\frac{1}{2} \left(\frac{1}{\epsilon_{UV}}+\ln \frac{\mu^2_{UV} }{(2p^z)^2}\right)\right. \right.\notag\\
&\left.\left.-\frac{1}{2}\ln^2 ((p^z)^2 \bm{b}_T^2 e^{2\gamma_E})+\frac{5}{2}\ln ((p^z)^2 \bm{b}_T^2 e^{2\gamma_E})-4\right]\right\}\,.
\end{align}
The Wilson line self energy diagram (Fig.~\ref{oneloop2}(c)) yields
\begin{align}
x \tilde{h}_{1T}^{\perp g}(x,\bm{b}_T^2,p^z)|_{\ref{oneloop2} (c)}=\frac{\alpha_s C_A}{2\pi}\delta(1-x)\left(\frac{3}{\epsilon_{UV}}+3 \ln \frac{\mu_{UV}^2 \bm{b}_T^2 e^{2\gamma_E}}{4}+2+\frac{2\pi L}{\left|\bm{b}_T\right|}\right)\,,
\end{align}
while Figs.~\ref{oneloop2}(d) and~\ref{oneloop2}(e) give vanishing contributions.

The factor in Eq.~(\ref{quasiZE}) and the renormalization factor in the $\overline{\mathrm{MS}}$ scheme, which were previously employed for the unpolarized and helicity gluon quasi-TMDPDFs $\tilde{f}_{1}^g$ and $\tilde{g}_{1L}^g$ in Ref.~\cite{Zhu:2022bja}, are:
\begin{align}
Z_E\left(2L, \left|\bm{b}_T\right|,\mu_{UV}\right)=&1+\frac{2 \alpha_s C_A}{\pi}\left(\frac{1}{\epsilon_{UV}}+\ln \frac{\mu_{UV}^2 \bm{b}_T^2 e^{2\gamma_E}}{4} +1+\frac{\pi L}{\left|\bm{b}_T\right|}\right)\,,\\
\tilde{Z}_1\left(\mu,\epsilon\right)=&1-\frac{\alpha_s }{4\pi}\left(\beta_0-2C_A\right)\frac{1}{\epsilon_{UV}}\,.
\end{align}
The renormalized and subtracted gluon quasi-TMDPDF $\tilde{h}_{1T}^{\perp g}(x,\bm{b}_T^2,\mu,p^z)$ is then obtained as
\begin{align}
&x \tilde{h}_{1T,sub}^{\perp g}(x,\bm{b}_T^2,\mu,p^z)\notag\\
=&\tilde{Z}_1\left(\mu,\epsilon\right) \frac{x \tilde{h}_{1T}^{\perp g,(0)}(x,\bm{b}_T^2,p^z)+x \tilde{h}_{1T}^{\perp g}(x,\bm{b}_T^2,p^z)}{\sqrt{Z_E\left(2L, \left|\bm{b}_T\right|,\mu_{UV}\right)}}\notag\\
=&\delta(1-x) -\frac{\alpha_s}{2\pi} \left(\frac{1}{\epsilon_{IR}}+\ln \frac{\mu^2 \bm{b}_T^2 e^{2 \gamma_E}}{4}\right)\cdot x \left[C_A \frac{2x}{(1-x)_+} + \frac{\beta_0}{2 } \delta (1-x)\right]\notag\\
&+\frac{\alpha_s C_A}{2\pi} \left[\ln \frac{\mu^2 \bm{b}_T^2 e^{2\gamma_E}}{4} \left(\frac{\beta_0}{2C_A}-1\right)-\frac{1}{2}\ln^2 \left(\left(p^z\right)^2 \bm{b}_T^2 e^{2\gamma_E}\right)+2 \ln \left(\left(p^z\right)^2 \bm{b}_T^2 e^{2\gamma_E}\right)  -4 \right] \delta(1-x)\,,
\end{align}
The relations $x\Delta \tilde{H}_{T,\text{sub}}^{g}(x,\bm{b}_T^2,\mu,p^z) = \kappa \, x\tilde{h}_{1T,\text{sub}}^{\perp g}(x,\bm{b}_T^2,\mu,p^z)$ and $x \tilde{h}_{1T,\text{sub}}^{g}(x,\bm{b}_T^2,\mu,p^z) = -2 \, x\tilde{h}_{1T,\text{sub}}^{\perp g}(x,\bm{b}_T^2,\mu,p^z)$ remain valid after renormalization and subtraction.

An important observation is that the differences among all gluon TMDPDFs and quasi-TMDPDFs originate solely from their splitting kernels in the infrared (IR) region. The distinction between the quasi-TMDPDFs and the standard TMDPDFs stems from differences in their ultraviolet (UV) structure, reflecting the fundamental difference between working with a finite $P^z$ (Euclidean) versus an infinite $P^+$ (light-cone) momentum. At one-loop order, the ratio between the renormalized quasi-TMDPDFs $\tilde{F}$ and the standard TMDPDFs $F$ for $\{F\} = \{f_{1}, g_{1L}, h_{1T}^{\perp}, h_{1T}, \Delta H_T\}$ is given by:
\begin{align}
\frac{\tilde{F}_{\overline{\mathrm{MS}}}^g \left(x,\bm{b}_T^2 ,\mu,\xi_z\right)}{F_{\overline{\mathrm{MS}}}^g \left(x,\bm{b}_T^2,\mu,\xi \right)}=&1+\frac{\alpha_s C_A}{2\pi}\left[-\ln \frac{\mu^2 \bm{b}_T^2 e^{2\gamma_E}}{4} \ln \frac{\xi_z}{\xi}+\ln \frac{\mu^2 \bm{b}_T^2 e^{2\gamma_E}}{4}\right.\notag\\
&\left.-\frac{1}{2} \ln^2 \frac{\xi_z}{\mu^2} +2 \ln \frac{\xi_z}{\mu^2}+\frac{\pi^2}{12}-4\right]\,,\label{ratio}
\end{align}
where $\xi_z=\left(2xP^z\right)^2$. Notably, the proportionality factors $\kappa$~$\left(\mathrm{or~-2}\right)$ cancel in the ratio. 

The matching relation for the gluon quasi-TMDPDFs is given by~\cite{Zhu:2022bja}
\begin{align}
\tilde{F}^{g}_{\overline{\mathrm{MS}}}\left(x,\bm{b}_T^2 ,\mu,\xi_z\right)S_r^{\frac{1}{2}}\left(\bm{b}_T^2,\mu\right)=H_F\left(\frac{\xi_z}{\mu^2}\right) e^{\ln \frac{\xi_z}{\xi} K\left(\bm{b}_T^2,\mu\right)}  F_{\overline{\mathrm{MS}}}^g \left(x,\bm{b}_T^2,\mu,\xi \right)\,,
\end{align}
where $S_r\left(\bm{b}_T^2,\mu\right)$ denotes the reduced soft function and $K\left(\bm{b}_T^2,\mu\right)$ is the Collins-Soper kernel. Eq.~(\ref{ratio}) implies that all these gluon quasi-TMDPDFs share the same matching coefficient at one-loop level
\begin{align}
H_F\left(\frac{\xi_z}{\mu^2}\right)=1+\frac{\alpha_s C_A}{2\pi}\left(-\frac{1}{2} \ln^2 \frac{\xi_z}{\mu^2} +2 \ln \frac{\xi_z}{\mu^2}+\frac{\pi^2}{12}-4\right)\,.
\end{align}

This universal matching coefficient for $f_1^g$, $g_{1L}^g$, $h_{1T}^{\perp g}$, $h_{1T}^g$, and $\Delta H_T^g$ stems from the identical UV structure difference between their quasi and standard counterparts. Although our explicit derivation employs the $\overline{\mathrm{MS}}$ scheme, the ratio in Eq.~(\ref{ratio}) is UV-finite and thus scheme-independent. Consequently, while the specific functional form of the hard function $H_F$ may vary under different renormalization schemes, the ratio of quasi-TMDPDFs to TMDPDFs—and therefore the physical matching relation remains invariant. The gluon TMDPDFs $f_{1T}^{\perp g}$, $g_{1T}^g$, $h_1^{\perp g}$, and $h_{1L}^{\perp g}$ exhibit a simpler structure at one-loop order, and we anticipate that future work will extend these results to two-loop accuracy. 

\section{Conclusions}\label{section4}

In this work, we have performed a perturbative calculation of the leading-twist gluon TMDPDFs and their quasi-distribution counterparts at one-loop order. By adopting a matrix formulation of the gluon-gluon correlator $\Phi_{\Lambda \Lambda\prime}^{ij}(x,\bm{k}_T^2;S)$ in the combined gluon $\otimes$ hadron spin space, utilizing circular polarization states, we systematically derived the one-loop expressions for both the standard TMDPDFs and quasi TMDPDFs. The dependence on the gluon transverse momentum was incorporated by working in a frame where the two components of the transverse momentum $\bm{k}_T^i$ are the same, and the sensitivity to the orientation of the transverse spin vector $\bm{S}_T$ was analyzed for the two representative cases:  $\bm{S}_T=\left(S_T^1,0\right)$ and $\bm{S}_T=\left(0,S_T^2\right)$.

Our calculations reveal several significant results. 
Among the eight leading-twist gluon TMDPDFs, $f_{1T}^{\perp g}$, $h_1^{\perp g}$, $h_{1T}^{g}$, and $h_{1T}^{\perp g}$ receive non-zero one-loop contributions, while $g_{1T}^{g}$ and $h_{1L}^{\perp g}$ are found to vanish at this order. A key finding is that despite their distinct IR structures—governed by different gluon-gluon splitting kernels—the distributions $f_1^g$, $g_{1L}^g$, $h_{1T}^{\perp g}$, $h_{1T}^g$, and the combined quantity $\Delta H_T^g$ yield an identical ratio between their quasi and light-cone counterparts. 
This universality originates from the fact that the difference between the quasi-TMDPDFs and the standard TMDPDFs resides in their UV structure, which is common to this set of distributions. Consequently, they share a universal matching coefficient at one-loop order.

These results establish the essential perturbative foundation for extracting this complete set of gluon TMDPDFs from first principles via lattice QCD simulations within the LaMET framework. 
The extension of LaMET to encompass all leading-twist gluon TMDPDFs, as presented here, represents a crucial step toward a comprehensive and unified determination of nucleon three-dimensional structure from lattice QCD. 
We look forward to future work extending these calculations to higher perturbative orders and to their practical implementation in lattice QCD computations. 

\section*{Acknowledgements}
X.X thanks Shuai Zhao for the very useful discussion on the loop integral of the wilson line for quasi-TMDPDFs. This work is partially supported by the National Natural Science Foundation of China under grant number 12150013. X. X is also supported by the SEU Innovation Capability Enhancement Plan for Doctoral Students under grant number CXJH$\_$SEU 25138.

\end{document}